\documentclass[12pt]{article}

\def\lsim{\mathrel{\rlap {\raise.5ex\hbox{$ < $}}
{\lower.5ex\hbox{$\sim$}}}}
\def\gsim{\mathrel{\rlap {\raise.5ex\hbox{$ > $}}
{\lower.5ex\hbox{$\sim$}}}}
\newcommand{\be}{\begin{eqnarray}}
\newcommand{\ee}{\end{eqnarray}}

\begin{document}

\begin{titlepage}

\begin{centering}

\hfill hep-ph/0104195\\

\vspace{1 in}
{\bf {A NOTE ON BRANE-COSMOLOGY} }\\
\vspace{2 cm}
{A. KEHAGIAS$^{1}$
 and K. TAMVAKIS$^{2}$}\\
\vskip 1cm
{$^1 $\it{Physics Department, National Technical University\\
15 780 Zografou, Athens,  GREECE}\\
\vskip 1cm
{$^2$\it {Physics Department, University of Ioannina\\
45110 Ioannina, GREECE}}}\\

\vspace{1.5cm}
{\bf Abstract}\\
\end{centering}
\vspace{.1in} We derive a new class of time-dependent solutions
for the Randall-Sundrum model by patching together isometries
broken by the brane. Solutions generated by generalized
boosts along the fifth dimension are associated with localized gravity and lead to an effective
Friedman equation on
 the brane with
a scale factor exhibiting power law or exponential behaviour. The effective
 energy-density on the brane
depends linearly on the brane tension.\vfill

\vspace{2cm}
\begin{flushleft}

April 2001
\end{flushleft}
\hrule width 6.7cm \vskip.1mm{\small \small}
 \end{titlepage}

The large gap between the electroweak scale and the gravitational Planck
 scale has motivated the quest of a theory with a higher dimensional spacetime in which our world corresponds to
a four-dimensional hypersurface. Although the idea that the world might correspond
 to a topological defect embedded in a higher dimensional spacetime is not new
 \cite{RS}, string theory provides many examples of such
hypersurfaces or {\textit{branes}} and, therefore, strong motivation for their exploration. In a class of such higher dimensional models
Standard Model Physics is confined on the brane while gravity propagates in the bulk.
 Large compact extra dimensions can lower the
fundamental gravitational scale down to the TeV range\cite{D1}. The extra space however does not have to be
 compact but just highly curved. Such is the Randall-Sundrum spacetime \cite{RS1}\cite{RS2} characterized by a graviton which, although
depending on the transverse coordinate, it is sharply localized on the brane.
The RS metric $ds_5^2=e^{-2\kappa|y|}g_{\mu\nu}dx^{\mu}dx^{\nu}+dy^2$
corresponds to gluing together two regions of the $AdS_5$ space
($y>0$ and $y<0$) such
that to posse a $Z_2$ symmetry. Smooth generalizations of such a geometry can be constructed with the help of a scalar field that gives rise to
the brane as a kink-like soliton while, in addition to gravitons,
other relevant degrees of freedom can be localized on the
brane\cite{KT1}.  An additional important possibility that seems to be
open in brane models is the possibility
of explaining the smallness of the observed cosmological constant.
Branes allow for an interplay between higher dimensional and
four-dimensional cosmological constant contributions. Such
{\textit{self-tuning}} behaviour \cite{Arkani-Hamed:2000eg}-\cite{Shafi:2001uf}
arises also in the case that the brane
 is realized as a scalar soliton\cite{KT2}.
An interesting as well as important issue is whether higher dimensional models lead to a cosmological evolution
compatible with standard Big Bang cosmology or extensions of it like inflationary models. The cosmological implications of
higher dimensional models have been studied by a number of authors
\cite{Lukas:1999yy}-\cite{jj}. A general
feature of brane cosmology is that in the Friedman-like equations on the brane the energy density of the brane, naivly
corresponding to the brane tension, appears quadratically \cite{B1} in
contrast to the Friedman equations of the standard cosmology  where it
appears linearly. In the Randall-Sundrum framework however,
this quadratic contribution in the right hand side of the Friedman equation is compensated by the linear bulk energy density coming from a bulk cosmological constant. Exact compensation gives the well known Randall-Sundrum
static flat solution. Dynamical evolution is possible in the form of exponential expansion when the above two contributions
do not cancel, leaving an effective four-dimensional energy density.

In the present article we investigate the time evolution on the brane in the framework of the Randall-Sundrum $AdS_5$ spacetime looking for time-dependent solutions of Einstein's
equations in the full five-dimensional space. We consider Lorentz transformations as a prototype of isometries
 that are broken by the presence of the brane and replace them with their $Z_2$-invariant analogues. The metric generated by such a transformation satisfies Einstein's equations with an effective energy-momentum tensor consisting of the contributions of the brane and a bulk cosmological constant. The corresponding cosmological evolution on the brane is that of exponential expansion. Next we consider a generalization of the above metric that leads to a general time-dependence of the four-dimensional scale factor that includes standard time evolution like $a(\tau)\propto \tau^{1/2}$. These solutions exist for specific time-dependent bulk energy-densities. In all cases the four-dimensional scale factor satisfies an effective Friedman equation that
 features the effective
energy-density on the brane calculated by summing the bulk energy-density contribution and the brane tension. The general quadratic dependence on the brane tension is replaced by a linear one due to the relation between the Randall-Sundrum curvature parameter
and the brane tension $\kappa=\xi/24M_5^3$ necessary for the existence of solutions.

Let us consider the five-dimensional Anti-de Sitter space $AdS_5$
with metric
\begin{equation}
ds^2=e^{-2\kappa
y}(-dt^2+dx_{\bot}^2+dx_{||}^2)=\frac{1}{\kappa^2x_{||}^2}d\sigma^2
\label{met}
\end{equation}
where the  notation
\begin{equation}x_{||}\equiv \frac{e^{\kappa y}}
{\kappa}\,\,,\,\,\,\,\,\,\,dx_{\bot}^2=\delta_{ij}dx^idx^j\, , ~~~~ i,j=1,2,3
\end{equation}
has been introduced. $AdS_5$ is maximally symmetric and its
Riemann tensor satisfies
$$
R_{\mu\nu\kappa\lambda}=\kappa\left(g_{\mu\kappa}g_{\nu\lambda}-g_{\nu\kappa}g_{\mu\lambda}
\right)\, , ~~~~~\mu\,...=0,...,4
$$
The metric $d\sigma^2$ and the $x_{||}=$const. slices of the full space are
 invariant under the {\textit{boosts}}
\begin{equation}t'=\frac{1}{\sqrt{1-\beta^2}}(t+\beta x_{||})\end{equation}
\begin{equation}x_{||}'=\frac{1}{\sqrt{1-\beta^2}}( x_{||}+\beta t)
\end{equation}
\begin{equation}{x_{\bot}'}^i=x_{\bot}^i\end{equation}
The total metric (\ref{met}) is then written in the new boosted coordinates as
\begin{equation}{ds}^2=\frac{1}{{\kappa^2x_{||}'}^2}
(-dt'^2+dx_{\bot}'^2+dx_{||}'^2)
=\frac{1-\beta^2}{{\kappa^2(x_{||}+\beta t)}^2}(-dt^2+dx_{\bot}^2+dx_{||}^2)
\end{equation}
In terms of the original coordinates we have, up to a constant
\begin{equation}ds^2=\frac{1}{(e^{\kappa y}+\beta t)^2}
\left(-dt^2+dx_1^2+dx_2^2+dx_3^2\right)+\frac{e^{2\kappa y}}
{(e^{\kappa y}+\beta t)^2}dy^2 \label{mmm}
\end{equation}
This metric still describes $AdS_5$ just in another coordinate
system, i.e. the boosted one. As a result,
we would  the same physics as the original metric
if no discontinuities exist, since the boosts are just a set of
general coordinate transformations. In the Randall-Sundrum model however,
where a brane sits at $y=0$, we can have the
above boosts for $y>0$ and their analogues
for $y<0$ such that no global coordinate transformation to exists
and describing different physics.
Introducing conformal time
$ e^{-\beta \tau}d\tau=-dt$ and
imposing the $Z_2$ symmetry $y\to -y$ we can set the  metric (\ref{mmm})
in the form
\be
ds^2=-\frac{e^{-2\beta\tau}d\tau^2}{(e^{\kappa |y|}+e^{-\beta \tau}-1)^2}
+\frac{\delta_{ij}dx^idx^j}{(e^{\kappa |y|}+e^{-\beta \tau}-1)^2}
+\frac{e^{2\kappa |y|}dy^2}{(e^{\kappa |y|}+e^{-\beta \tau}-1)^2}
\label{MM}
\ee
When the parameter $\beta$ vanishes, this metric is the static Randall-Sundrum
metric
$$ds_{RS}^2=e^{-2\kappa|y|}
g_{\mu\nu}dx^{\mu}dx^{\nu}+dy^2$$
Dynamical solutions with the same static limit has been constructed also in 
\cite{Grojean:2000az},\cite{Binetruy:2000wn}.
It is clear that if the {\textit{right}} ($y>0$) or {\textit{left}} ($y<0$)
 metric given (\ref{MM}) was extended to the whole spacetime,
it would provide a solution of Einstein's equations with a cosmological 
constant.
The existence of the brane, mathematically
represented by the presence of the absolute value $|y|$,
will induce extra $\delta(y)$ and constant
terms coming from $|y|^{\prime\prime}$ and $(|y|')^2$. Such terms are exactly 
the terms expected from a bulk cosmological constant
and a 3-brane term in the action. Thus, the metric (8) should be compatible 
with the action
\begin{equation}
{\cal{S}}=\int d^5x\sqrt{-g}\left\{2M_5^3R-\Lambda_B\right\}-
\int d^5x\sqrt{-\det(g_{\mu\nu})}\,\xi \delta(y)
\end{equation}
which leads to Einstein's equations
\begin{equation}G_{MN}=R_{MN}-\frac{1}{2}g_{MN}R=-g_{MN}\frac{\Lambda_B}
{4M_5^3}-g_{\mu\nu}\delta_M^{\mu}\delta_N^{\nu}\frac{\xi}{4M_5^3}
\frac{\delta(y)}{b}\equiv \frac{1}{4M_5^3}T_{MN}\end{equation}
The two parameters appearing in these equations, namely the 
{\textit{brane tension}} $\xi$ and
 the {\textit{bulk cosmological constant}} $\Lambda_B$ should be matched with 
the two parameters of the metric $\kappa$ and $\beta$.

Using the metric ansatz
\begin{equation}ds^2=-n^2(\tau, y)d\tau^2+a^2(\tau,y)\gamma_{ij}dx^idx^j+
b^2(\tau,y)dy^2
\end{equation}
 with $\gamma_{ij}$ the metric of a three-dimensional maximally symmetric 
space\footnote{$\gamma_{rr}=(1-kr^2)^{-1}$, $\gamma_{\theta\theta}=r^2$, 
$\gamma_{\phi\phi}=r^2\sin^2\theta$}, we can compute components of  
Einstein's tensor $G_{AB}$. They are\cite{B2}\cite{K}
\begin{equation}G_{00}=3\left\{\frac{\dot{a}}{a}\left(\frac{\dot{a}}{a}+
\frac{\dot{b}}{b}\right)-\frac{n^2}{b^2}\left(\frac{a^{\prime\prime}}{a}
+\frac{a'}{a}\left(\frac{a'}{a}-\frac{b'}{b}\right)\right)+k
\frac{n^2}{a^2}\right\}\end{equation}
\begin{equation}G_{ij}=\frac{a^2}{b^2}\gamma_{ij}\left\{\frac{a'}{a}
\left(\frac{a'}{a}+2\frac{n'}{n}\right)-
\frac{b'}{b}\left(\frac{n'}{n}+2\frac{a'}{a}\right)+
2\frac{a^{\prime \prime}}{a}+\frac{n^{\prime \prime}}{n}\right\}$$
$$+\frac{a^2}{n^2}\gamma_{ij}\left\{\frac{\dot{a}}{a}
\left(-\frac{\dot{a}}{a}+2\frac{\dot{n}}{n}\right)
-2\frac{\ddot{a}}{a}+\frac{\dot{b}}{b}\left(-2\frac{\dot{a}}{a}+
\frac{\dot{n}}{n}\right)-\frac{\ddot{b}}{b}\right\}-k\gamma_{ij}\end{equation}
\begin{equation}G_{05}=3\left(\frac{n'}{n}\frac{\dot{a}}{a}+
\frac{a'}{a}\frac{\dot{b}}{b}-\frac{\dot{a}'}{a}\right)\end{equation}
\begin{equation}G_{55}=3\left\{\frac{a'}{a}\left(\frac{a'}{a}+
\frac{n'}{n}\right)-\frac{b^2}{n^2}\left(\frac{\dot{a}}{a}
\left(\frac{\dot{a}}{a}
-\frac{\dot{n}}{n}\right)+\frac{\ddot{a}}{a}\right)-k\frac{b^2}{a^2}\right\}
\end{equation}
Substituting the boosted expressions (9), (10) and (11) of the metric 
components, we can easily see that they are indeed a solution to 
Einstein's equations for the case $k=0$. We also determine the solution 
parameters in terms of the parameters of the action $\Lambda_B$ and $\xi$.
Their relations are

\begin{equation}\xi=24M_5^3\kappa\end{equation}
\begin{equation}\Lambda_B=-24M_5^3(\kappa^2-\beta^2)\end{equation}
Notice that for $\beta=0$, when the metric coincides with the static 
Randall-Sundrum metric, these relations are the
well known relations that relate the brane tension to the curvature 
parameter and express the fine-tuning between the bulk cosmological 
constant and the brane tension in order to have a flat solution. 
This last relation however in the
 case of non-vanishing $\beta$ should not be interpreted as a fine-tuning 
since there is a continuous range of values of
$\kappa^2+\frac{\Lambda_B}{24M_5^3}\geq 0$ corresponding to solutions.

The four-dimensional metric corresponding to our solution is
\begin{equation}ds_4^2=-d\tau^2+a_0^2(\tau)d\vec{x}^2\end{equation}
with
\begin{equation}a_0(\tau)=e^{\beta \tau}\end{equation}
This exponential expansion can be described with an 
{\textit{effective Friedman equation}}
\begin{equation}
\left(\frac{\dot{a}_0}{a_0}\right)^2=
\frac{8\pi G}{3}\rho_{eff}\end{equation}
This can be easily seen considering that\footnote{The four-dimensional 
Planck-mass is determined to be $M_P^2=M_5^3/\kappa$.
Using standard time we have $ds_4^2=a^2(t,0)(-dt^2+d\vec{x}^2)=
g_{\mu\nu}(t)dx^{\mu}dx^{\nu}$ and $\sqrt{-G}=\sqrt{-g}\,b(t,y)
(a(t,y)/a(t,0))^4$. Since,
 $$R[G]=\frac{a^2(t,y)}{a^2(t,0)}R^{(4)}[g]+\dots$$
 performing the $y$-integration,
we obtain
$${\cal{S}}=2M_5^3\int d^4x\int dy\,\sqrt{-G}R=2M_5^3\int d^4x
\sqrt{-g}a^{-2}(t,0)\left\{\int dya^2(t,y)b(t,y)\right\}R^{(4)}+\dots$$

$$=\frac{2M_5^3}{\kappa}\int d^4x\sqrt{-g}R^{(4)}[g]+\dots$$
from which we read off
$$M_P^2=M_5^3/\kappa$$} $8\pi G=\frac{1}{4M_P^2}=\frac{\kappa}{4M_5^3}$. 
Then, the effective energy density should satisfy
\begin{equation}\left(\frac{\dot{a}_0}{a_0}\right)^2=\beta^2=
\frac{\kappa}{12M_5^3}\rho_{eff}\end{equation}
In view of the parameter relations (18) and (19), the effective 
four-dimensional energy density is
\begin{equation}\rho_{eff}=\frac{1}{2}(\frac{\Lambda_B}{\kappa}+\xi)=\frac{1}{2}(\rho_B^{(4)}+\rho_b^{(4)})\end{equation}
being the average of the bulk contribution to the four-dimensional energy density\footnote{Note that $\kappa^{-1}$ is the effective
size of the fifth dimension.} $\Lambda_B\times \kappa^{-1}$ and the naive
four dimensional energy density $\xi$. Note that in the static Randall-Sundrum case these two terms cancel to give an exactly
vanishing $\rho_{eff}$. It is  worth comparing the above effective Friedman equation with the analogous general equation\cite{B1} derived for
any constant bulk energy-momentum
\begin{equation}\left(\frac{\dot{a}_0}{a_0}\right)^2=\frac{1}{24M_5^3}\rho_B+\frac{1}{(24M_5^3)^2}\rho_b^2-\frac{k}{a_0^2}+\frac{{\cal{C}}}{a_0^4}\end{equation}
The two equations coincide for $k={\cal{C}}=0$ due to the relation $\rho_b=\xi=24M_5^3\kappa$ which transforms the quadratic dependence on the brane tension into a linear one, namely
$$\left(\frac{\dot{a}_0}{a_0}\right)^2=\frac{\kappa}{24M_5^3}\left(\rho_B^{(4)}+\rho_b^{(4)}\right)=$$
$$\frac{\kappa}{24M_5^3}\left(\frac{\rho_B^{(5)}}{\kappa}+\frac{(\rho_b^{(4)})^2}{24M_5^3\kappa}\right)=
\frac{1}{24M_5^3}\left(\rho_B^{(5)}+\frac{(\rho_b^{(4)})^2}{24M_5^3}\right)$$
It seems, therefore, misleading to interpret the above equation as indicating a quadratic dependence on the brane energy density.

The next step in the quest for time-dependent solutions compatible with standard
 cosmology would be to replace the exponential in the metric ansatz with a general function of time $a_0(\tau)$. Introducing
the generalized trial solutions

\begin{equation}a(\tau,y)=\frac{a_0(\tau)}{a_0(\tau)(e^{\kappa|y|}-1)+1}\end{equation}
\begin{equation}n(\tau,y)=\frac{1}{a_0(\tau)(e^{\kappa|y|}-1)+1}=\frac{a}{a_0}\end{equation}
\begin{equation}b(\tau,y)=\frac{a_0(\tau)e^{\kappa|y|}}{a_0(\tau)(e^{\kappa|y|}-1)+1}=e^{\kappa|y|}a\end{equation}
we obtain, for $k=0$, the Einstein tensors

\begin{equation}G_{05}=0\end{equation}
\begin{equation}G_{00}
=g_{00}\left\{6\kappa^2-6\left(\frac{\dot{a}_0}{a_0}\right)^2-6\kappa\frac{\delta(y)}{b_0}\right\}\end{equation}
\begin{equation}G_{55}=g_{55}\left\{6\kappa^2-6\left(\frac{\dot{a}_0}{a_0}\right)^2-\frac{3}{n(\tau,y)}\left(\frac{\ddot{a}_0}{a_0}-\left(\frac{\dot{a}_0}{a_0}\right)^2\right)
\right\}\end{equation}
\begin{equation}G_{ij}=g_{ij}\left\{6\kappa^2-6\left(\frac{\dot{a}_0}{a_0}\right)^2-6\kappa\frac{\delta(y)}{b_0}-\frac{3}{n(\tau,y)}\left(\frac{\ddot{a}_0}{a_0}-\left(\frac{\dot{a}_0}{a_0}\right)^2\right)
\right\}\end{equation}
These, through Einstein's equations, can be matched with a conserved energy-momentum tensor
\begin{equation}T_N^M=Diag\left(-\rho,\,p,\,p,\,p,\,p_T\right)\end{equation}
with components
\begin{equation}\rho=24M_5^3\left\{\kappa \frac{\delta(y)}{b_0}-\kappa^2+\left(\frac{\dot{a}_0}{a_0}\right)^2\right\}\end{equation}
\begin{equation}p=-\rho-\frac{12M_5^3}{n(\tau,y)}\left\{\frac{\ddot{a}_0}{a_0}-
\left(\frac{\dot{a}_0}{a_0}\right)^2\right\}\end{equation}
$p_T$ equals to the bulk part of $p$. Note that the bulk energy density is only time-dependent while the bulk pressure
has also a space-dependence. Notice that there is no time-dependence introduced on the brane tension which is again $\xi=24M_5^3\kappa$. It is easy to see that the content of the previous section can be easily recovered from the above energy-momentum
tensor which becomes constant.

Let as now introduce a power law time-dependence in the scale factor
\begin{equation}a_0(\tau)=C\tau^{\gamma}\end{equation}
The 5D metric then turns out to be
\begin{equation}
ds^2=-\frac{d\tau^2}{C^2\tau^{2\gamma}(e^{\kappa|y|}-1)+1}
+\frac{C^2\tau^{2\gamma}\delta_{ij}dx^idx^j}
{C^2\tau^{2\gamma}(e^{\kappa|y|}-1)+1}
+\frac{C^2\tau^{2\gamma}e^{\kappa|y|}dy^2}{C^2\tau^{2\gamma}
(e^{\kappa|y|}-1)+1}
\end{equation}
whereas the
corresponding  densities are
\begin{equation}\rho(\tau)=24M_5^3\left\{\kappa \frac{\delta(y)}{b_0}-\kappa^2+\frac{\gamma^2}{\tau^2}\right\}\end{equation}
\begin{equation}p(\tau,y)=-\rho(\tau)+12M_5^3\frac{\gamma}{\tau^2}\left\{\frac{1}{n(\tau,y)}-2\gamma\right\}\end{equation}
$p_T$ is given by the bulk-part of $p$.

A physical explanation for the spatial dependence of the pressure is given by the fact that the force along the fifth dimension takes the form
\begin{equation}f(\tau,y)\equiv -\frac{\partial p}{\partial y}= 12M_5^3C\gamma \tau^{\gamma-2}\kappa sign(y)e^{\kappa |y|}\end{equation}
representing a force directed towards the brane and increasing with distance from it. A bulk pressure density of this
form is apparently required to sustain the brane at its given position.

The scale factor resulting from such an energy-momentum density satisfies an effective Friedman equation
\begin{equation}\left(\frac{\dot{a}_0}{a_0}\right)^2=\frac{\gamma^2}{\tau^2}=\frac{\kappa}{12M_5^3}\rho_{eff}\end{equation}
Thus,
\begin{equation}\rho_{eff}=\frac{\rho_B(\tau)}{2\kappa}+12M_5^3\kappa=\frac{1}{2}\left(\frac{\rho_B(\tau)}{\kappa}+\xi\right)=\frac{1}{2}(\rho_B^{(4)}(\tau)+\rho_b^{(4)})\end{equation}
in terms of the bulk part of the five-dimensional energy density $\rho_B$ and the brane tension. Thus, again we have the same averaging formula as in
the time-independent case and there is no quadratic dependence on the brane tension. Time-dependent perturbations on the brane tension will not modify the linear dependence. This is clear from
$$\left(\frac{\dot{a}_0}{a_0}\right)^2=\frac{1}{24M_5^3}\left(\rho_B^{(5)}+\frac{(\rho_b^{(4)})^2}{24M_5^3}\right)=$$
$$\frac{1}{24M_5^3}\left(\rho_B^{(5)}+\frac{(\rho_b^{(4)}+\delta\xi(\tau))^2}{24M_5^3}\right)\sim \frac{\kappa}{24M_5^3}\left(\rho_B^{(4)}+\rho_b^{(4)}+2\delta\xi(\tau)\right)+O(\delta\xi^2)$$

From the above analysis it is clear that both cases of inflation
($a_0(\tau)\propto e^{\beta \tau}$)
and standard cosmological expansion ($a_0(\tau)\propto \tau^{1/2}$) could be
described by a common suitable bulk energy density. Indeed, the bulk energy density
\begin{equation}
\rho_B(\tau)=24M_5^3\left\{-\kappa^2+\frac{\beta^2 \kappa^{-1}}{1+\beta^2\gamma^{-2}
\tau^2}\right\}
\end{equation}
corresponds to a scale factor
\begin{equation}a_0(\tau)\propto e^{\gamma
\sinh^{-1}\left({\beta \over
\gamma}\tau\right)}\end{equation}
Using that $\sinh^{-1}(x)\sim x $ for $x<<1$ and
$\sinh^{-1}(x)\sim \ln 2x$ for $x>>1$, we get
an early exponential expansion and a late power-law one
\begin{equation}a_0(\tau)\propto \left\{\begin{array}{cc}
e^{\beta \tau}\,&\, \tau\rightarrow 0\\
\tau^{\gamma}\,&\,\tau\rightarrow \infty
\end{array}\right.\end{equation}
A bulk energy density of this sort would describe a temporary inflationary phase succeded by a phase of standard
power law expansion.

Summarizing, we have studied a class of time-dependent solutions of Einstein's
equations in the framework of a five-dimensional spacetime corresponding to the standard Randall-Sundrum $AdS$ metric in
the static case. We considered Lorentz transformations as a prototype of isometries
 that are broken by the presence of the brane and replace them with their $Z_2$-invariant analogues. The metric generated by such a transformation satisfies Einstein's equations with an effective energy-momentum tensor consisting of the contributions of the brane and a bulk cosmological constant. The corresponding cosmological evolution on the brane is that of exponential expansion. Next we studied a generalization of the above metric corresponding to a general time-dependence of the four-dimensional scale factor that includes standard time evolution like $a(\tau)\propto \tau^{1/2}$. These solutions exist for specific time-dependent bulk energy-densities. In all cases the four-dimensional scale factor satisfies an effective Friedman equation that
 features the effective
energy-density on the brane calculated by summing the bulk energy-density contribution and the brane tension. The general quadratic dependence on the brane tension is replaced by a linear one due to the relation between the Randall-Sundrum curvature parameter
and the brane tension.

\newpage

\noindent
{\textbf{Acknowledgements}}

K.T. would like to thank Panagiota Kanti and Alexios Polychronakos for enlightening discussions.
 This work is partially
supported by the  RTN contracts HPRN-CT-2000-00122 and
HPRN-CT-2000-00131 and the $\Gamma\Gamma$ET grant E$\Lambda$/71.

\end{document}